\documentclass[prd,twocolumn,showpacs,floatfix,amsmath,nofootinbib,floatfix]{revtex4}
\usepackage[utf8x]{inputenc}
\usepackage{dcolumn,booktabs,bm,multirow}
\usepackage{longtable,lscape}
\usepackage{txfonts}
\usepackage{overpic}
\usepackage{amssymb}
\usepackage{array}
\usepackage{indentfirst}
\usepackage{feynmf}   %{feynmp}
\usepackage{slashed}  %for Feynman symbols
\usepackage{cases}
\usepackage{color}
\usepackage{multirow}
\usepackage{setspace}
\usepackage{graphicx}
\usepackage[colorlinks,
            citecolor=green,
            anchorcolor=red,
            menucolor=red,
            linkcolor=red,
            filecolor=red,
            runcolor=red,
            urlcolor=blue,
            frenchlinks=red]{hyperref}
\usepackage{epstopdf}

\begin{document}

\title{Production of the $Z_b^{(\prime)}$ states from the $\Upsilon(5S,6S)$ decays}

\author{Qi Wu$^{1}$}\email{wuq@seu.edu.cn}
\author{Dian-Yong Chen$^{1}$\footnote{Corresponding author}}\email{chendy@seu.edu.cn}
\author{Feng-Kun Guo$^{2,3}$}\email{fkguo@itp.ac.cn}

\affiliation{
$^1$School of Physics, Southeast University, Nanjing 210094, China\\
$^2$CAS Key Laboratory of Theoretical Physics, Institute of Theoretical Physics, Chinese Academy of Sciences, Beijing 100190, China\\
$^3$School of Physical Sciences, University of Chinese Academy of Sciences, Beijing 100049, China
}

\begin{abstract}
In the present work, we investigate the production mechanism of the $Z_b(10610)$ and $Z_b(10650)$ states from the $\Upsilon(5S,6S)$ decays. Two types of bottom-meson loops are discussed. We show that the loop contributions with all intermediate states being the $S$-wave ground state bottom mesons are negligible, while the loops with one bottom meson being the broad $B_0^\ast$ or $B_1^\prime$ resonance could provide the dominant contributions to the $\Upsilon(5S) \to Z_b^{(\prime)} \pi$. It is found that such a mechanism is not suppressed by the large width of the $B_0/B_1'$ resonance. In addition, we also estimate the branching ratios for the $\Upsilon(6S) \to Z_b^{(\prime)} \pi$ which could be  tested by future precise measurements at Belle-II.
\end{abstract}

\date{\today}
\maketitle

%%%%%%%%%%%%%%%%%%%%%%%%%%%%%%%%%%
\section{Introduction}
\label{sec:introduction}

In the past decade, a growing number of new hadron states have been observed, which are dubbed as $XYZ$ states in the heavy quarkonium mass regions (for recent reviews, we refer to Refs.~\cite{Chen:2016qju,Hosaka:2016pey,Lebed:2016hpi,Esposito:2016noz,Guo:2017jvc,Ali:2017jda,Olsen:2017bmm,Karliner:2017qhf,Yuan:2018inv,Kou:2018nap}). Unlike the prosperity of charmoniumlike states, in the bottom sector, only two such bottomoniumlike states have been observed, which are the $Z_b(10610)$ and the $Z_b(10650)$, to be denoted as $Z_b$ and $Z_b^{\prime}$, respectively.
These two bottomoniumlike states were firstly reported in the $\Upsilon(nS) \pi^\pm,\ (n=1,2,3) $ and $h_b(mP) \pi^\pm,\  (m=1,2)$ invariant mass distributions of the dipion decays of the $\Upsilon(10860)$\footnote{The $\Upsilon(10860)$ and $\Upsilon(11020)$ will be loosely called $\Upsilon(5S)$ and $\Upsilon(6S)$, respectively, in the paper.} by the Belle Collaboration in 2011~\cite{Collaboration:2011gja,Belle:2011aa}.
Later on, the neutral partners of $Z_b$ and $Z_b^\prime$ were also discovered~\cite{Krokovny:2013mgx}. The analyses of the charged pion angular distributions suggest that the quantum numbers of both $Z_b$ and $Z_b^\prime$ be $I^G(J^P)=1^+(1^+)$~\cite{Garmash:2014dhx}. Besides the hidden-bottom decay modes, both $Z_b$ and $Z_b^\prime$ have also been observed in the open-bottom decay channels of the $\Upsilon(5S)$~\cite{Adachi:2012cx,Garmash:2015rfd}. Moreover, the Belle Collaboration reported their measurements of the transitions $\Upsilon(6S) \to \pi^+ \pi^- h_b(mP)$~\cite{Abdesselam:2015zza}, and the measured $h_b(mP) \pi, \ (m=1,2)$ invariant mass spectra indicated that the decays $\Upsilon(6S)\to h_b(mP) \pi^+ \pi^-$ proceed entirely via the intermediate $Z_b$ and $Z_b^\prime$ states.

Since the observed $Z_b$ and $Z_b^\prime$ are isospin triplets and their masses are in the bottomonium mass region, they contain at least four valence quarks ($b\bar b q\bar q$ with $q=u,d$) if they are hadronic resonances. They were thus proposed to be tetraquark states~\cite{Guo:2011gu,Cui:2011fj,Ali:2011ug,Wang:2013zra,Ali:2014dva,Patel:2016otd}. There are two salient features of the $Z_b^{(\prime)}$ states: (1) Although their masses are very close to the $B^\ast \bar{B}$ and $B^\ast \bar{B}^\ast$ thresholds, respectively, they still decay dominantly into the open-bottom final states~\cite{Garmash:2015rfd}. (2) They decay into the heavy quark spin-triplet $\Upsilon\pi$ and spin-singlet $h_b\pi$ final states with similar rates~\cite{Belle:2011aa}. Moreover, their quantum numbers allow them to couple to a pair of bottom and anti-bottom ground state mesons in $S$-waves.
These features suggest to consider the $Z_b$ and $Z_b^\prime$ as the deuteronlike molecular states composed of $B^\ast \bar{B}$ and $B^\ast \bar{B}^\ast$, respectively~\cite{Bondar:2011ev,Sun:2011uh,Mehen:2011yh,Cleven:2011gp,Li:2012wf,Yang:2011rp,Zhang:2011jja,Wang:2013daa, Wang:2014gwa,Dong:2012hc,Ohkoda:2013cea,Li:2012as,Cleven:2013sq,Li:2012uc, Li:2014pfa,Dias:2014pva,Xiao:2017uve}. In this scenario, the $b\bar{b}$ pairs in both $Z_b$ and $Z_b^\prime$ are mixtures of a spin-triplet and a spin-singlet, and thus the observations with similar rates of the $Z_b^{(\prime)}$ in both final states containing the spin-triplet $\Upsilon(1S,2S,3S)$ and spin-singlet $h_b(1P,2P)$ can be naturally understood~\cite{Bondar:2011ev}.

The $Z_b$ and $Z_b'$ masses given in the original Belle measurements are above the $B\bar B^*$ and $B^*\bar B^*$ thresholds, respectively~\cite{Collaboration:2011gja,Belle:2011aa}. However, it is subtle to precisely determine the masses due to the very nearby $S$-wave thresholds. Detailed analyses of the $Z_b$ and $Z_b'$ line shapes have been made in past years by considering the strong coupling of the $Z_b^{(\prime)}$ to the open-bottom channels~\cite{Cleven:2011gp,Mehen:2013mva,Hanhart:2015cua,Guo:2016bjq,Wang:2018jlv}, and the most advanced analysis shows that the $Z_b$ pole is slightly below the $B\bar B^*$ threshold while the $Z_b'$ pole is slightly above the $B^*\bar B^*$ threshold~\cite{Wang:2018jlv}.

\begin{table*}
\centering
\caption{The experimental measurements of the related branching ratios, where $\mathcal{B}_{\Upsilon(5S)}^{}=\mathcal{B}(\Upsilon(5S) \to (b\bar{b}) \pi^+ \pi^-)$, $\mathcal{B}_{Z_b^{(\prime)}}=\mathcal{B}(Z_b^{(\prime)} \to (b\bar{b}) \pi)$ and $f_{Z_b^{(\prime)}} $ is the fractions of individual quasi-two-body channels contributions to $\Upsilon(5S) \to Z_b^{(\prime)^\pm} \pi^\mp \to (b\bar{b}) \pi^+ \pi^-$, where $(b\bar{b})$ could be $\Upsilon(nS),\ (n=1,2,3)$ and $h_b(mP),\ (m=1,2)$. The branching ratios $\Upsilon(5S) \to Z_b^+ \pi^-$ and $\Upsilon(5S)\to Z_b^{\prime +} \pi^-$ are estimated by the measured data.
 \label{Table:Exp}}
 \renewcommand\arraystretch{1.35}
\begin{tabular}{p{0.8cm}<{\centering}p{2.5cm}<{\centering}p{2cm}<{\centering}p{2cm}<{\centering}p{2cm}<{\centering}p{2cm}<{\centering}|p{2cm}<{\centering}p{2cm}<{\centering}}
\toprule[1pt]
 & $\mathcal{B}_{\Upsilon(5S)}\ (10^{-3})$~\cite{Tanabashi:2018oca} & ${f}_{Z_b}\ (\%)$ \cite{Adachi:2012cx} & ${f}_{Z_b^\prime} \ (\%)$ \cite{Adachi:2012cx}& $\mathcal{B}_{Z_b} \ (\%)$ \cite{Garmash:2015rfd} & $\mathcal{B}_{Z_b^\prime}\ (\%) $ \cite{Garmash:2015rfd} & $\mathcal{B}(\Upsilon(5S) \to Z_b^+ \pi^-)\  (\%)$ & $\mathcal{B}(\Upsilon(5S) \to Z_b^{\prime +} \pi^-)\  (\%)$\\
 \midrule[1pt]
$\Upsilon(1S)$& $5.3 \pm 0.6$ & $2.54^{+0.86+0.13}_{-0.51-0.55}$& $1.04^{+0.65+0.07}_{-0.31-0.12}$ &$0.54^{+0.16+0.11}_{-0.13-0.08}$ & $0.17^{+0.07+0.03}_{-0.06-0.02}$ & $1.25^{+0.63}_{-0.52}$ & $1.62^{+1.26}_{-0.82}$\\
$\Upsilon(2S)$& $7.8 \pm 1.3$ & $19.6^{+3.5+1.9}_{-3.1-0.6}$& $5.77^{+1.44+0.27}_{-0.96-1.56}$ & $3.62^{+0.76+0.79}_{-0.59-0.53}$ & $1.39^{+0.48+0.34}_{-0.38-0.23}$ & $2.11^{+0.84}_{-0.67}$ & $1.62^{+0.84}_{-0.67}$\\
$\Upsilon(3S)$& $4.8^{+1.0}_{-1.7}$ & $26.8^{+6.6}_{-3.9} \pm 1.5$ & $11.0^{+4.2}_{-2.3} \pm 0.7$ & $2.15^{+0.55+0.60}_{-0.42-0.43}$ & $1.63^{+0.53+0.39}_{-0.42-0.28} $ & $2.99^{+1.80}_{-1.42}$& $1.62^{+1.13}_{-0.84}$\\
$h_b(1P)$ & $3.5^{+1.0}_{-1.3}$ & $42.3 ^{+9.5 +6.7}_{-12.7-0.8}$ & $60.2 ^{+10.3+4.1}_{-12.7-3.8}$& $3.45^{+0.87+0.86}_{-0.71-0.63}$ & $8.41^{+2.43+1.49}_{-2.12-1.06}$ & $2.15^{+1.14}_{-1.18}$ & $1.25^{+0.60}_{-0.73}$\\
$h_b(2P)$ & $5.7^{+1.7}_{-2.1}$ & $35.2^{+15.6+0.1}_{-0.4-13.4}$ & $64.8 ^{+15.2+6.7}_{-11.4-15.5}$ & $4.67^{+1.24+1.18}_{-1.00-0.89}$ & $14.7^{+3.2+2.8}_{-2.8-2.3}$&
$2.15^{+1.39}_{-1.41}$ & $1.26^{+0.61}_{-0.60}$\\
\bottomrule[1pt]
\end{tabular}
\end{table*}

The above literature focuses mostly on the resonance parameters and the decay behaviors of $Z_b$ and $Z_b^\prime$. However, the productions of these two bottomoniumlike states also have interesting issues.  From the experimental side, the decay patterns of $Z_b^{(\prime)}$ have been measured~\cite{Collaboration:2011gja, Belle:2011aa,Adachi:2012cx,Garmash:2015rfd}, and in addition, the Belle Collaboration also reported the fractions of individual quasi-two-body contributions to $\Upsilon(5S) \to Z_b^{(\prime)\pm} \pi^\mp \to (b\bar{b}) \pi^+ \pi^-$, where $(b\bar{b})$ denotes $\Upsilon(nS),\ (n=1,2,3)$ or $h_b(mP),\ (m=1,2)$.
All the related experimental data are listed in Table~\ref{Table:Exp}. One can relate the branching ratios of the $\Upsilon(5S) \to Z_b^{(\prime)} \pi$ to the measured fractions by $\mathcal{B}(\Upsilon(5S) \to Z_b^{(\prime)} \pi)= f_{Z_b^{(\prime)}} \mathcal{B}_{\Upsilon(5S)}/\mathcal{B}_{Z_b^{(\prime)}}^{}$. With the experimental data listed in Table~\ref{Table:Exp}, the branching ratios of the $\Upsilon(5S) \to Z_b^{(\prime)\pm} \pi^\mp$ can be approximately estimated, which are also listed in Table~\ref{Table:Exp}.
One finds that the branching ratios from different channels are consistent with each other within errors\footnote{Notice that the Belle analysis in Ref.~\cite{Garmash:2015rfd}, where no values for $f_{Z_b^{(\prime)}}$ are given, presents an update of that in Ref.~\cite{Adachi:2012cx}, and thus there is inconsistency in using them simultaneously. This is why these branching fractions in the last column of the table do not agree with each other exactly.} and are of the order of $10^{-2}$.
Given that the sum of the non-open-bottom branching fractions of the $\Upsilon(5S)$ is only $\left(3.8^{+5.0}_{-0.5}\right)\%$ as given in the 2018 Review of Particle Physics by the Particle Data Group (PDG)~\cite{Tanabashi:2018oca}, such values are surprisingly large. It is thus interesting to understand the reasons behind.

{
}

The production of the $Z_b^{(\prime)}$ states in the $\Upsilon(5S)$ decays have been modeled by considering either direct $\Upsilon Z_b\pi$ couplings or through intermediate ground state bottom-meson loops~\cite{Chen:2011zv, Cleven:2011gp,Mehen:2013mva,Hanhart:2015cua,Guo:2016bjq, Wang:2018jlv}. The latter mechanism is shown in Fig.~\ref{fig:feyn-Zb-1}.
As noticed in Ref.~\cite{Cleven:2011gp} and will be briefly analyzed in Sec.~\ref{sec:formula}, such loops are expected to contribute little.
For the production of the $Z_b^{(\prime)}$ states in $\Upsilon(6S)$ decays, since the $\Upsilon(6S)$ is very close to the thresholds of $B_1(5271) \bar{B}$, it was pointed out in Refs.~\cite{Wang:2013hga,Bondar:2016pox} that triangle singularities (see the reviews~\cite{Guo:2017jvc,Guo:2017wzr} and references therein) could be important to enhance the production rates.
However, the narrow $B_1(5721)$ is mainly a meson with $s_\ell^P=3/2^+$, where $P$ denotes the parity and $s_\ell$ is the total angular momentum of the light quark system which becomes a good quantum number in the heavy quark limit~\cite{ManoharWise}, and it has been shown that the $S$-wave production of a pair of $3/2^+$ and $1/2^-$ (i.e., ground state $S$-wave heavy mesons) mesons in $e^+e^-$ collisions is suppressed in the heavy quark limit~\cite{Li:2013yka}. Thus, a mixing between $3/2^+$ and $1/2^+$ axial-vector bottom mesons, though suppressed in the heavy quark limit as well, is introduced in Ref.~\cite{Bondar:2016pox}.

In this paper, we will point out the importance of bottom-meson loops with one bottom meson being the $s_\ell^P=1/2^+$ state which has a large width. As will be shown here, the large width will enhance, instead of weaken, the contribution from such loops.  Arguments based on power counting in a nonrelativistic effective field theory (NREFT)~\cite{Guo:2009wr,Guo:2010ak,Guo:2017jvc} will be presented in Sec.~\ref{sec:formula}, and the numerical results showing explicitly the importance will be given in Sec.~\ref{sec:results}. Section~\ref{sec:summary} is devoted to a short summary.

\section{Meson loop contributions to $\Upsilon(5S, 6S) \to Z_b^{(\prime)} \pi$   }
\label{sec:formula}

\begin{figure}[ht]
\begin{tabular}{ccc}
  \centering
 \includegraphics[width=2.75cm]{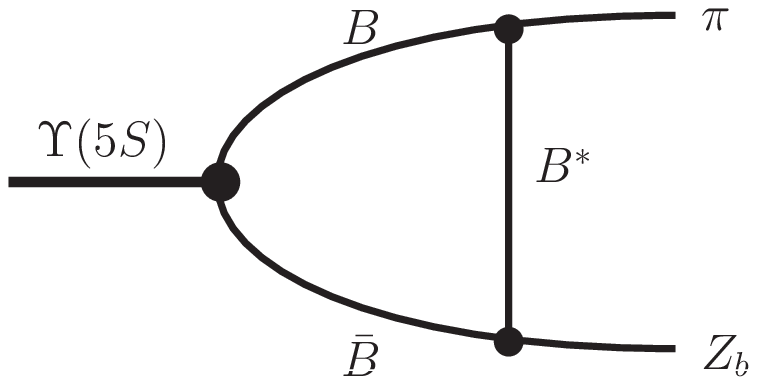}&
 \includegraphics[width=2.75cm]{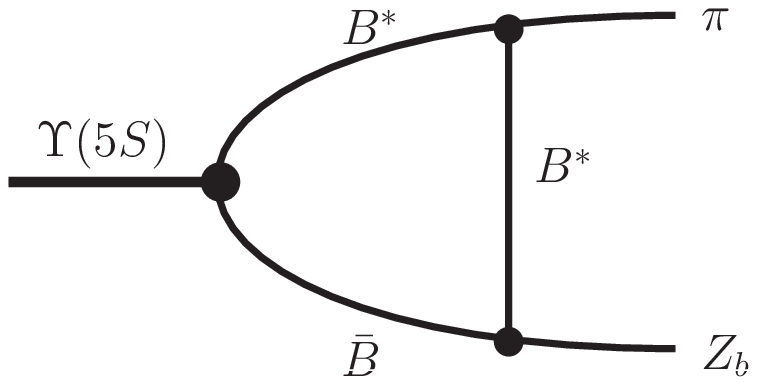}&
 \includegraphics[width=2.75cm]{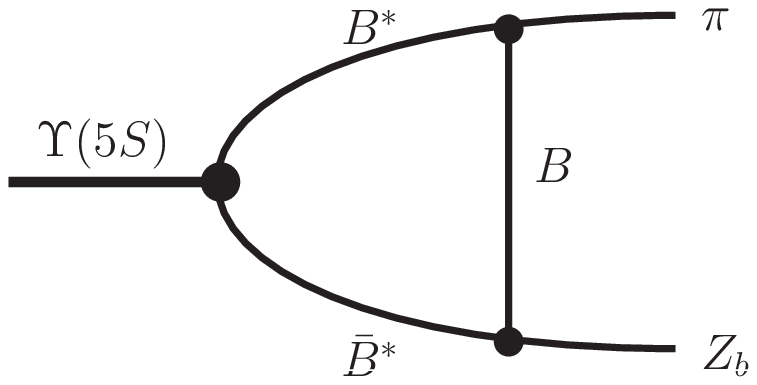}\\
 \\
 $(a)$ & $(b)$ &$(c)$\\
 \\
 \includegraphics[width=2.75cm]{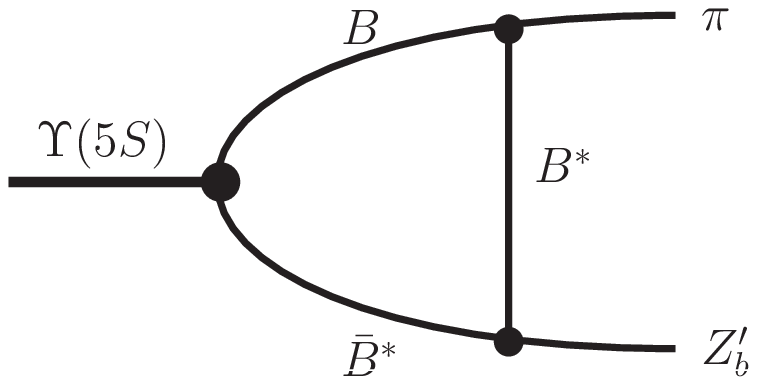}&
 \includegraphics[width=2.75cm]{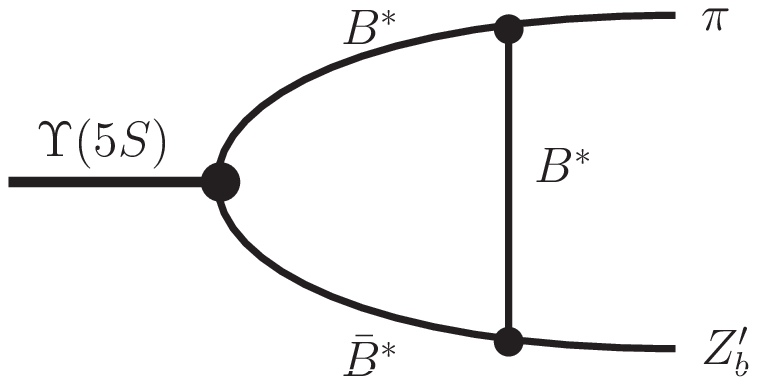}&\\
 \\
 $(d)$ & $(e)$&\\
 \end{tabular}
  \caption{Hadron-level diagrams for the $\Upsilon(5S) \to Z^{(\prime)+}_b \pi^-$ processes via the $B^{(\ast)}\bar{B}^{(\ast)}$ intermediate heavy meson loops.}\label{fig:feyn-Zb-1}
\end{figure}

\begin{figure}[ht]
\begin{tabular}{ccc}
  \centering
  \includegraphics[width=2.75cm]{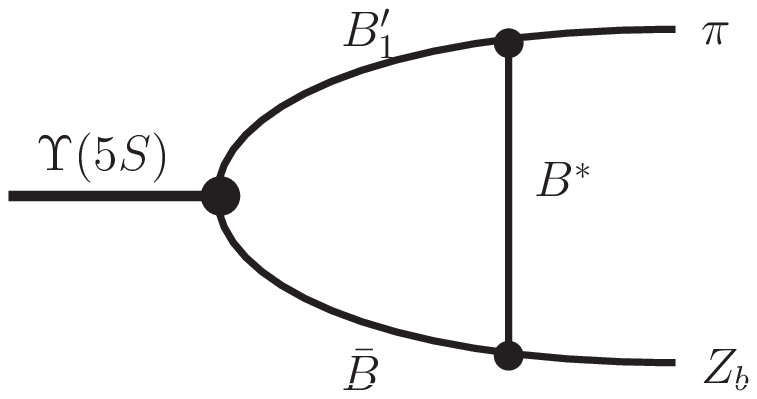}&
  \includegraphics[width=2.75cm]{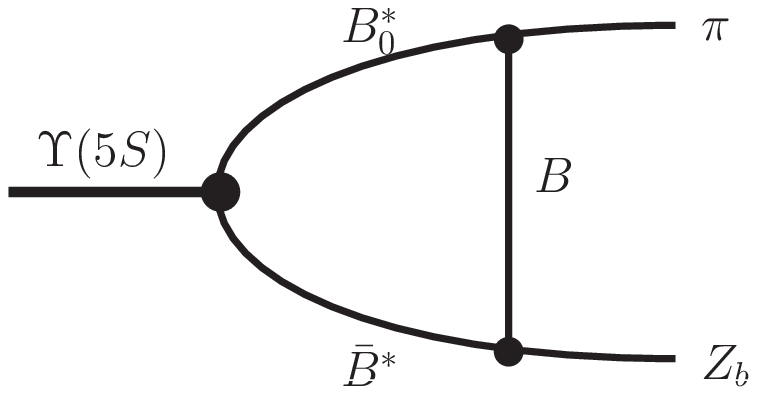}&
  \includegraphics[width=2.75cm]{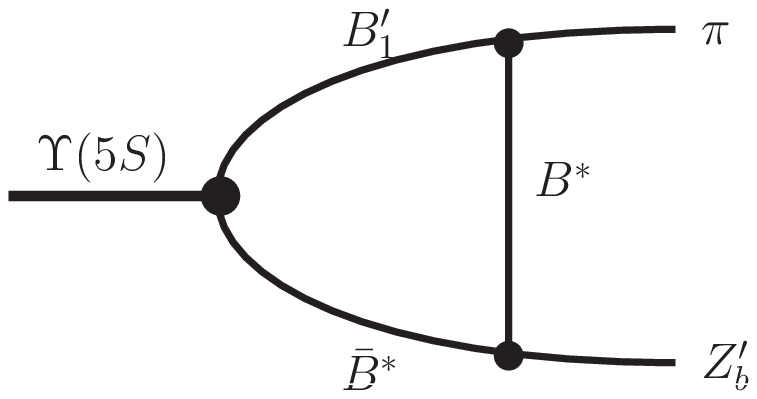}\\
  \\
 $(a)$ & $(b)$ & $(c)$\\
 \end{tabular}
  \caption{Hadron-level diagrams for the $\Upsilon(5S) \to Z^{(\prime)+}_b \pi^-$ processes via the $B^{\prime0}_1 \bar{B}^{(\ast)0} /B^{\ast0}_0 \bar{B}^{\ast0} +c.c.$ intermediate heavy meson loops.}\label{fig:feyn-Zb-3}
\end{figure}

As for a bottomonium above the open-bottom threshold, it dominantly decays into a pair of bottom mesons, and the bottom meson pair can couple to the final states via exchanging a proper bottom meson. Such a kind of mechanism may play a primary role in understanding some decay modes of higher heavy quarkonium or heavy-quarkonium-like states~\cite{Mehen:2011tp, Zhang:2018eeo, Guo:2010ak,Huang:2017kkg, Chen:2014ccr, Chen:2013cpa, Chen:2011jp, Chen:2011qx, Xiao:2017uve}. In particular, taking $\Upsilon(5S)\to Z_b^{(\prime)} \pi$ as an example, the initial $\Upsilon(5S)$ dominantly decays into a pair of $S$-wave bottom mesons, i.e., $B \bar{B}$, $B^\ast \bar{B}+c.c.$ and $B^\ast \bar{B}^\ast$.
By exchanging a bottom meson, these bottom meson pairs can transit into the $Z_b^{(\prime)} \pi$.
The corresponding diagrams contributed to the $\Upsilon(5S) \to Z_b^{(\prime)} \pi$ are presented in Fig.~\ref{fig:feyn-Zb-1}. However, as will be shown later, since both the $\Upsilon(5S)$ and the pion couple to the $B^{(\ast)} \bar{B}^{(\ast)}$ in $P$-waves, these diagrams are highly suppressed.

In addition, it should be noticed that the mass of the $\Upsilon(5S)$ is located in the vicinity of the $B^{(\ast)} B_1^\prime$  and $B^\ast B_0^{\ast}$ thresholds, where $B_0^*$ and $B_1'$ refer to the lowest $s_\ell^P=1/2^+$ bottom mesons. Thus, the $\Upsilon(5S) \to Z_b^{(\prime)} \pi$ processes can proceed via the mechanism shown in Fig.~\ref{fig:feyn-Zb-3}.  In this meson loop,  all the involved vertices, $\Upsilon(5S) B_1^\prime B$, $B_1^\prime B^\ast \pi$, $\Upsilon(5S) B_0^\ast B^\ast$ and $B_0^\ast B \pi$ are in $S$-waves,  leading to an enhancement in comparison with the mechanism in Fig.~\ref{fig:feyn-Zb-1} as will be shown below.
In the following, we analyze these two kinds of mechanisms by the NREFT power counting rule~\cite{Guo:2009wr,Guo:2010ak,Guo:2017jvc}.

\subsection{$B^{(\ast)}\bar{B}^{(\ast)}+c.c.$ meson loops}

In NREFT, one of the key quantities of the power counting rule is the typical velocity $v\ll 1$ of the nonrelativistic intermediate mesons. The momentum and nonrelativistic energy count as $v$ and $v^2$, respectively. The integral measure scales as $v^5$, and the heavy meson propagator counts as $1/v^2$. The $S$-wave vertices are independent on the velocity.  While $P$-wave vertices are much more complicated, it scales either as $v$ or the external momentum~\cite{Guo:2017jvc}.

As presented in Fig.~\ref{fig:feyn-Zb-1}, the initial bottomonium $\Upsilon(5S)$ connects to the final  $Z_b^{(\prime)} \pi $ via $B^{(\ast)} \bar B^{(\ast)}$ loops. In these diagrams, both the $\Upsilon(5S) B^{(\ast)}\bar{B}^{(\ast)} $ and $B^{\ast} {B}^{(\ast)} \pi$ vertices have a $P$-wave coupling, while the $Z^{(\prime)}_b$ couples to the $B^{(\ast)}\bar B^{(\ast)}$ in $S$-waves.
As discussed in Refs.~\cite{Guo:2012tg,Guo:2017jvc}, there are two momentum scales in the nonrelativistic triangle diagrams, corresponding to the two momenta of the bottom mesons connected to the initial and final heavy particles.
They are given by $\sqrt{|c_1|}$ and $\sqrt{|c_2|}$ with $c_1$ and $c_2$ defined in Eq.~\eqref{eq:c1c2} in Appendix~\ref{appendix-A}. Accordingly, one can define two velocities for the intermediate mesons, which are, $v_1=\sqrt{|c_1|}/(2\mu_{12})$ and $v_2=\sqrt{|c_2-a|}/(2\mu_{23})$, where $\mu_{ij}$ and $a$ are also defined in Eq.~\eqref{eq:c1c2}.
Here, the velocity in the NREFT power counting corresponds to the average of these two velocities~\cite{Guo:2012tg,Guo:2017jvc}, i.e., $v=(v_1+v_2)/2$. For the diagrams in Fig.~\ref{fig:feyn-Zb-1},  we denote the velocity as $v_A$ and one has $v_A=0.12 \sim 0.14$, which indicates that the corresponding amplitudes could be analyzed in a nonrelativistic framework.

Both the $\Upsilon(5S) B^{(\ast)} \bar B^{(\ast)}$ and $B^{\ast} B^{(\ast)} \pi$ vertices are $P$-wave couplings. The latter coupling introduces a factor of $\vec{q}$ to the amplitudes, where $\vec{q}$ is the pion momentum. The former vertex brings an internal momentum, which turns into the external momentum $\vec{q}$ after performing the loop integrals. As a result, the amplitude from the mechanism in Fig.~\ref{fig:feyn-Zb-1} scales as~\cite{Guo:2010zk,Guo:2010ak,Guo:2017jvc}
\begin{equation}
\mathcal{A}_A \sim N_A\, \frac{{v_A^5}\vec{q}^{\,2}}{(v_A^2)^3m_B^2} =  N_A \frac{\vec{q}^{\,2}}{v_A^{}m_B^2},\label{Eq:A1}
\end{equation}
where $N_A$ collects all constant factors including, e.g., the coupling constants, the loop geometrical factor and the normalization factors, and a factor of $1/m_B^2$ with $m_B$ being the bottom meson mass is introduced to balance the dimension of $\vec q^{\,2}$. In fact, the amplitude here is similar to that for $\psi'\to h_c\pi$, except for the latter breaking isospin symmetry, and has been shown to be highly suppressed when the pion momentum is much smaller than the intermediate heavy meson mass as detailed in Ref.~\cite{Guo:2010zk}.

\subsection{$B^\prime_1 \bar{B}^{(\ast)}/B^{\ast}_0 \bar{B}^{\ast} +c.c.$ meson loops}

Besides the the $B^{(\ast)} \bar{B}^{(\ast)}$ meson loops, the initial $\Upsilon(5S)$ and final $Z_b^{(\prime)} \pi$ can also be bridged by the $B^\prime_1 \bar{B}^{(\ast)}/B^{\ast}_0 \bar{B}^{\ast} +c.c$ meson loops, as presented in Fig.~\ref{fig:feyn-Zb-3}.  In these kinds of meson loops, all of the involved interaction vertices are $S$-wave coupling. We denote the velocity as $v_B$, and the corresponding amplitude scales as
\begin{equation}
\mathcal{A}_B \sim N_B\, \frac{v_B^5  E_\pi^{}}{(v_B^2)^3m_B^{}} =  N_B \frac{E_\pi^{}}{v_B^{}m_B^{}}, \label{Eq:A2}
\end{equation}
where $N_B$ collects all the constant factors, $E_\pi$ comes from the pionic $S$-wave coupling, and a factor of $1/m_B$ is introduced to balance the dimension of $E_\pi$. From Eq.~(\ref{Eq:A2}), one can find that $\mathcal{A}_2$ is proportional to $1/v_B$, which indicates that the amplitude is greatly enhanced for a small velocity.
To date, the $B_1^\prime$ and $B^\ast_0$ have not been discovered yet. We adopt the values $m_{B_1^{\prime}} =5584$~MeV and $m_{B_0^{\ast}} =5535$~MeV~\cite{Du:2017zvv}, which are predicted using the heavy quark flavor symmetry in a framework which can describe both the lattice~\cite{Liu:2012zya,Moir:2016srx} and experimental data~\cite{Aaij:2016fma} for the $D\pi$ $S$-wave systems~\cite{Albaladejo:2016lbb,Du:2017zvv}. Numerically, $v_B=0.05\sim 0.07$, which is about 2 times smaller than $v_A$ in the $B^{(\ast)} \bar{B}^{(\ast)} +c.c$ loops. Notice that here the large widths $\sim 200$~MeV of the $1/2^+$ mesons have not been taken into account. Considering the width using a complex mass $m-i\,\Gamma/2$, one sees that the width effect in the power counting is to increase the absolute value for $v_B$ to roughly in the same ballpark as $v_A$. We will discuss their effect in the explicit calculations in Sec.~\ref{sec:results}.

With the amplitude scalings presented in Eqs.~(\ref{Eq:A1}) and (\ref{Eq:A2}), we roughly estimate the ratio of the contributions from the $B^\prime_1 \bar{B}^{(\ast)}/B^{\ast}_0 \bar{B}^{\ast} +c.c.$ loops and the $B^{(\ast)} \bar{B}^{(\ast)}+c.c.$ loops, which is
\begin{eqnarray}
\frac{\mathcal{A}_B}{\mathcal{A}_A}\sim \frac{m_B E_\pi v_A}{ \vec{q}^{\,2} v_B  } = \mathcal{O}(30),
\label{Eq:AmpR}
\end{eqnarray}
assuming $N_A\sim N_B$ which is reasonable as long as all the couplings take natural values.
This means that the contribution from the $B^\prime_1 \bar{B}^{(\ast)}/B^{\ast}_0 \bar{B}^{\ast} +c.c.$ meson loops should be much larger than that from the $S$-wave bottom mesons, and can potentially lead to a large rate for the $Z_b^{(\prime)}$ productions from the $\Upsilon(5S)$ decays.

\section{Explicit calculation of the bottom-meson loops}
\label{sec:results}

\subsection{Effective Lagrangian}

In this section, we  present a detailed calculation of these diagrams in Figs.~\ref{fig:feyn-Zb-1} and \ref{fig:feyn-Zb-3} in the NREFT framework, which is widely employed to study transitions between heavy quarkonium(-like) states~\cite{Guo:2009wr,Guo:2010zk,Guo:2010ak,Cleven:2011gp,Mehen:2011tp,Li:2013xia,Esposito:2014hsa,Mehen:2015efa,Wu:2016dws,Zhang:2018eeo}. To calculate diagrams presented in Figs.~\ref{fig:feyn-Zb-1} and \ref{fig:feyn-Zb-3}, we employ the effective Lagrangians constructed in the heavy quark limit.
In this limit, the $S$-wave heavy-light mesons form a spin multiplet $H=\{P,V\}$ with $s_\ell^P=1/2^-$, where $P$ and $V$ denote the pseudoscalar and vector heavy mesons, respectively. The $s_\ell^P=1/2^+$ states are collected in $S=\{P_0^\ast, P_1^\prime\}$ with $P_0^\ast$ and $P_1^\prime$ denoting the $B_0^*$ and $B_1'$ states, respectively. It is worthwhile to notice that we avoid to use ``$P$-wave mesons'' for these states as they could well be dynamically generated from the interaction between the $1/2^-$ states and the light pseudoscalar mesons (pions, kaons and $\eta$), see Ref.~\cite{Du:2017zvv} and references therein. Nevertheless, their quantum numbers are still  $s_\ell^P=1/2^+$ and form a spin multiplet. Using the two-component notation~\cite{Hu:2005gf}, the spin multiplets are given by
\begin{eqnarray}
H_{a}&=&\vec{V}_a\cdot\vec{\sigma}+P_a,\nonumber\\
S_{a}&=&\vec{P}_{1a}'\cdot\vec{\sigma}+ P_{0a}^*,
\end{eqnarray}
where $\vec\sigma$ denotes the Pauli matrices, and $a$ is the light-flavor index.
The fields for their charge conjugated mesons are
\begin{eqnarray}
\bar H_{a} &=&-\vec{\bar{V}}_a\cdot\vec{\sigma}+\bar{P}_a,\nonumber\\
\bar S_{a} &=&-\vec{\bar{P}}_{1a}\cdot\vec{\sigma}+\bar{V}_{0a}.
\end{eqnarray}
The field for the spin multiplet of the $S$-wave $\Upsilon$ and $\eta_b$ states is given by
\begin{eqnarray}
\Upsilon=\vec{\Upsilon}\cdot\vec{\sigma}+\eta_b.
\end{eqnarray}
%In the heavy meson rest frame where $v^\mu=(1,0,0,0)$,
The effective Lagrangian for the $S$-wave bottomonia coupled to a pair of $1/2^-$ bottom mesons is~\cite{Guo:2009wr}
\begin{equation}
\mathcal{L}=i\frac{g_1}{2}\ \mathrm{Tr}[\Upsilon^\dag {H}_{a}\vec{\sigma}\cdot{\stackrel{\leftrightarrow}{\partial}}\bar{H}_{a}]+\ \mathrm{H.c.}, \label{Eq:LHH}
\end{equation}
while the coupling between the $S$-wave bottomonia and a $1/2^-$-$1/2^+$ pair of bottom mesons is
\begin{equation}
\mathcal{L}=g_{2}\ \mathrm{Tr}[\Upsilon^\dag {S}_{a}\bar {H}_{a}+\Upsilon^\dag {H}_{a}\bar{S}_{a}] +\ \mathrm{H.c.} \label{Eq:LHS}
\end{equation}
We will use $g_2$ and $g_2'$ for the couplings for the $\Upsilon(5S)$ and $\Upsilon(6S)$, respectively.
Assuming that the $Z_b$ and $Z^{\prime}_{b}$ couple to $B\bar B^*$ and $B^*\bar B^*$, respectively~\cite{Bondar:2011ev}, the effective Lagrangian is given by~\cite{Cleven:2013sq}
\begin{eqnarray}
\mathcal{L}=z^{\prime}\varepsilon^{ijk}\bar{V}^{\dag i}Z^{\prime j}V^{\dag k}+z[\bar{V}^{\dag i}Z^{i}P^{\dag}-\bar{P}^{\dag}Z^{i}V^{\dag i}] + \ \mathrm{H.c.}, \label{Eq:LZ}
\end{eqnarray}
where $z$ and $z'$ are effective couplings.

The pionic couplings to heavy mesons are constrained by chiral symmetry. For the $S$-wave heavy mesons, the leading order Lagrangian in heavy meson chiral perturbation theory is given by~\cite{Wise:1992hn,Hu:2005gf}
\begin{eqnarray}
\mathcal{L}=-\frac{g}{2}\ \mathrm{Tr}[H^\dag_{a} H_{b}\vec{\sigma}\cdot\vec{u}_{ba}],\label{Eq:Lchi}
\end{eqnarray}
where the axial current is $\vec{u}=-\sqrt{2}\vec{\partial}\phi/F_\pi + \mathcal{O}(\phi^3)$. Here, $F_\pi$ the pion decay constant in the chiral limit, and
\[ \phi = \begin{pmatrix}
\pi^0/\sqrt{2} & \pi^+ \\
\pi^- & -\pi^0/\sqrt{2}
\end{pmatrix}
\]
collects the pion fields.
The leading order Lagrangian for the pions coupled to a pair of $s_\ell^P=1/2^+$ and $s_\ell^P=1/2^-$ heavy-light mesons is~\cite{Kilian:1992hq,Casalbuoni:1996pg}
\begin{eqnarray}
\mathcal{L}=i\frac{h}{2}\ \mathrm{Tr}[H^\dag_{a} S_{b} u^0_{ba}] + H.c.,\label{Eq:Lchi}
\end{eqnarray}
where $u^0=-\sqrt{2}\partial^0\phi/F_\pi+\mathcal{O}(\phi^3)$.

\subsection{Numerical results and discussion}

\renewcommand\arraystretch{1.25}
\begin{table*}[tb]
\centering
\caption{Branching ratios obtained calculated assuming only the mechanisms depicted in Figs.~\ref{fig:feyn-Zb-1} and \ref{fig:feyn-Zb-3}. Here $g_2$ and $g_2'$ take values in units of GeV$^{-1/2}$. \label{Table:BR}}\label{tab:state-2}
\centering
\begin{tabular}{p{2.5cm}<\centering p{3cm}<\centering p{3cm}<\centering p{3cm}<\centering p{3cm}<\centering}
\toprule[1pt]
&
$\mathcal{B}(\Upsilon(5S) \to Z_b^+ \pi^-)$ &
$\mathcal{B}(\Upsilon(5S) \to Z_b^{\prime +} \pi^-)$ &
$\mathcal{B}(\Upsilon(6S) \to Z_b^+ \pi^-)$ &
$\mathcal{B}(\Upsilon(6S) \to Z_b^{\prime +} \pi^-)$ \\
\midrule[1pt]
$B^{(\ast)} \bar{B}^{(\ast)}$ Loops &
$6.1 \times 10^{-4}$ &
$2.8 \times 10^{-4}$ &
$4.1 \times 10^{-4}$ &
$1.9 \times 10^{-4}$ \\
$B_1^\prime \bar{B}^{(\ast)}$ Loops &
$9.5 g_2^2 $ &
$3.2 g_2^2 $ &
$17.3 g_2^{\prime 2} $ &
$8.3 g_2^{\prime 2} $ \\
\bottomrule[1pt]
\end{tabular}
\end{table*}

Using the the measured branching fractions and widths of the $\Upsilon(5S,6S)$~\cite{Tanabashi:2018oca}, the coupling constant $g_1$ in Eq.~(\ref{Eq:LHH}) is estimated to be $0.1 \ \mathrm{GeV}^{-3/2}$ and $0.08 \ \mathrm{GeV}^{-3/2}$ for the $\Upsilon(5S)$\footnote{Here we neglect the heavy quark spin symmetry breaking effect discussed in Ref.~\cite{Mehen:2013mva}.} and $\Upsilon(6S)$, respectively.
From the effective Lagrangian in Eq. (\ref{Eq:LZ}), one gets the partial widths of $Z_b \to B^\ast \bar{B} +h.c$ and $Z_b^\prime \to B^\ast \bar{B}^\ast$ as
\begin{eqnarray}
\Gamma[Z^{+}_b\rightarrow B^{\ast+}\bar{B}^0+\bar{B}^{\ast0}B^+] &=& \frac{1}{4\pi}\frac{|\vec{q}|}{M_{Z_b}}|z|^2 M_B M_{B^\ast},\nonumber\\
\Gamma[Z^{\prime+}_b\rightarrow B^{\ast+}\bar{B}^{\ast0}]&=&\frac{1}{4\pi}\frac{|\vec{q}|}{M_{Z^\prime_b}}|z^\prime|^2  M^2_{B^\ast},
\end{eqnarray}
respectively. Here we take the PDG averages of the widths of the $Z_b^{(\prime)}$~\cite{Tanabashi:2018oca} and the measured branching ratios of the open bottom channels~\cite{Garmash:2015rfd} to get the values of the coupling constants, which are,
\begin{eqnarray}
z &=&(0.77\pm0.05)\ \mathrm{GeV}^{-1/2},\nonumber\\
z^{\prime}&=&(0.58\pm0.07) \ \mathrm{GeV}^{-1/2}.
\end{eqnarray}

The total widths of the $B^\prime_1$ and the $B^\ast_0$ are approximately saturated by the decays $B^\prime_1\rightarrow B^{\ast}\pi$,  and $B^\ast_0\rightarrow B\pi$, and their decay widths are
\begin{eqnarray}
\Gamma(B^{\prime0}_1\rightarrow B^{\ast} \pi)=\frac{3h^2}{8\pi F^2_\pi}\frac{m_{B^\ast}}{m_{B^{\prime}_1}}\left(m^2_\pi+|\vec{p}|^2\right)|\vec{p}|,\nonumber\\
\Gamma(B^{\ast0}_0\rightarrow B \pi)=\frac{3h^2}{8\pi F^2_\pi}\frac{m_B}{m_{B^{\ast}_0}}\left(m^2_\pi+|\vec{p}|^2\right)|\vec{p}|,\label{Eq:B1B0}
\end{eqnarray}
where we have multiplied the amplitude by a factor of $\sqrt{m_\text{ex}}$ for each external bottom meson to take into account the nonrelativistic normalization, with $m_\text{ex}$ the external bottom meson mass, and both $B^{(*)+}\pi^-$ and $B^{(*)0}\pi^0$ are considered. Using the central values of the resonances parameters in Ref.~\cite{Du:2017zvv}, which are $m_{B^\ast_0}=5535\ \mathrm{MeV},\ \Gamma_{B^\ast_0}=226\ \mathrm{MeV}, \ m_{B_1^\prime}=5584\ \mathrm{MeV},\ \Gamma_{B_1^\prime}=238\ \mathrm{MeV}$, we get $|h|\simeq1.1$. Similarly, the axial coupling $g$ is determined from the $D^{*+}\to D^0\pi^+$ decay width to be $|g|\simeq0.57$.

The large widths of the $B_1^\prime$ and the $B^{\ast}_0$ need to be taken into account in the calculations.
We introduce the width effects by approximating the spectral function of the broad $1/2^+$ bottom mesons using the Breit-Wigner (BW) parametrization.\footnote{In fact, the BW form is not a good parametrization for the line shapes of the broad $B_1^\prime$ and $B_0$ as discussed in Ref.~\cite{Du:2017zvv}. However, since here we are only interested in the effects caused by the widths, rather than the line shapes, of the $B_1^\prime$ and $B_0$, the BW form should suffice.} The explicit formula for the $B^{\ast}_0$ is
\begin{equation}
  \mathcal{M}_{B_0^*} = \frac1{W_{B_0^*}^{}} \int_{s_l}^{s_h} ds\, \rho_{B_0^*}^{}(s) \bar{\mathcal{M}}_{B_0^*}(s) \, ,
  \label{eq:sf}
\end{equation}
where $\bar{\mathcal{M}}_{B_0^*}(s)$ represents the  loop amplitude involving the $B_0^*$ calculated using $s$ as its mass squared, $s_l={(M_B+m_\pi)^2}$, $s_h$ is taken to be ${(M_{B^\ast_0}+\Gamma_{B_0^\ast})^2}$,  $\rho_{B_0^*}(s)$ is the $B^{\ast}_0$ spectral function
\begin{equation}
  \rho_{B_0^*}^{}(s) = \frac1{\pi} \text{Im} \frac{-1}{s - M_{B_0^*}^2 + i\, M_{B_0^*}^{} \Gamma_{B_0^*}^{}} \,,
  \label{eq:sfB0}
\end{equation}
and $W_{B_0^*}^{} = \int_{s_l}^{s_h} ds \rho_{B_0^*}^{}(s)$ is the normalization factor. The formula for the $B_1'$ is similar.

With the above coupling constants and the amplitudes in Eqs.~(\ref{Eq:Amp1}) and (\ref{Eq:Amp2}), we can compute different bottom-meson loop contributions to the $\Upsilon(5S) \to Z_b^{(\prime)} \pi$ decays. The obtained branching ratios considering only the  mechanisms depicted in Figs.~\ref{fig:feyn-Zb-1} and \ref{fig:feyn-Zb-3} are presented in Table~\ref{tab:state-2}.  By comparing with the branching ratios in Table~\ref{Table:Exp}, one finds that the contributions from the $B^{(\ast)} \bar{B}^{(\ast)} +c.c$ loops are two orders of magnitude smaller than the experimental data. This means that the $B^{(\ast)} \bar{B}^{(\ast)}$ meson loops can be neglected in the production of the $Z_b^{(\prime)}$. On the other hand,  as indicated in Eq.~(\ref{Eq:AmpR}), the amplitude resulted from the $B_1^\prime \bar{B}^{(\ast)}/B^{\ast}_0 \bar{B}^{\ast}$   loops is about 30 times larger than the one from the $B^{(\ast)} \bar{B}^{(\ast)} +c.c$ loops, which implies that the contribution to the partial widths from the former kind of loops is at least two orders of magnitude larger than that from the latter. Based on these two facts, one can conclude that
the $B_1^\prime \bar{B}^{(\ast)}/B^{\ast}_0 \bar{B}^{\ast}  +c.c$ loops could be the dominant production mechanism of $\Upsilon(5S) \to Z_b^{(\prime)} \pi$, though the value for the effective coupling constant $g_2$ is unknown.

\begin{figure}[tb]
  \centering
 \includegraphics[width=0.95\linewidth]{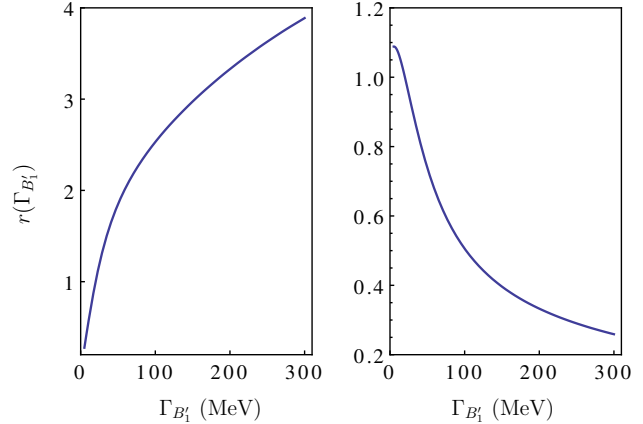}
  \caption{The $\Gamma_{B^\prime_1}$-dependence of $r(\Gamma_{B_1'})$ defined in Eq.~\eqref{eq:retio} with $h$ determined from $\Gamma_{B^\prime_1}$ (left) and with $h$ fixed to a constant value (right).}\label{fig:ratios}
\end{figure}

Broad resonances are rarely considered in the literature discussing meson loops.\footnote{In the analogous charm sector, a small contribution from the broad $D_1(2430)$ was introduced to provide an the description of the $e^+e^-\to D\bar D^*\pi$ process~\cite{Cleven:2013mka,Qin:2016spb,thesis}.} The main reason is that it is implicitly assumed that the large width entering the propagator of the broad resonance would highly suppress its contribution. Here we investigate the width effect quantitatively. Taking the $\Upsilon(5S)\to Z_b^{\prime+}\pi^-$ as an example,
we calculate its width $\Gamma_{(5S)}'$ as a function of the width of the $B_1'$. To make the width effect transparent, we define the following ratio
\begin{equation}
 r(\Gamma_{B_1'}) \equiv \frac{\Gamma_{(5S)}'(\Gamma_{B_1'})}{\Gamma_{(5S)}'(\Gamma_{B_1'}=20\,\text{MeV})},
\label{eq:retio}
\end{equation}
where the benchmark width 20~MeV is an arbitrarily chosen small width.
It is worthwhile to notice that the $B_1'$ width depends on the coupling $h$ defined in Eq.~\eqref{Eq:B1B0}, and the same coupling enters the $B_0^*B\pi$ and $B_1'B^*\pi$ vertices in Fig.~\ref{fig:feyn-Zb-3}.
Thus, while a large $h$ value---thus a large $B_1'$ width---suppresses the loop integral, it also provides an enhancement factor as $\Gamma_{(5S)}'$ from the mechanism in Fig.~\ref{fig:feyn-Zb-3} is explicitly proportional to $h^2$. Therefore, the $B_1'$ width effect depends on their competition. In the left panel of Fig.~\ref{fig:ratios}, the result of $r(\Gamma_{B_1'})$ is depicted, showing an enhancement instead of a suppression.
If we only consider the width effect in the $B_1'$ propagator with $h$ fixed to a constant value, the result is shown in the right panel of Fig.~\ref{fig:ratios} for a comparison.\footnote{Note that although this does not correspond to the physical situation at hand, it is relevant for the processes when the vertex in the triangle diagram does not give the dominant decay channel of the intermediate resonance.} In this case, one sees that the result using a width of about 200~MeV is about 30\% of that using a width of 20~MeV.

% \begin{figure}[ht]
% \setlength{\tabcolsep}{6mm}{
% \begin{tabular}{ccc}
%   \centering
%   % Requires \usepackage{graphicx}
%  \includegraphics[width=7cm]{width_effect_b}
%  \end{tabular}}
%   \caption{The $\Gamma_{B^\prime_1}$-dependence of $r(\Gamma_{B_1'})$ defined in Eq.~\ref{eq:retio} with $h$ fixed.}\label{fig:ratio b}
% \end{figure}

% \begin{figure}[ht]
% \setlength{\tabcolsep}{6mm}{
% \begin{tabular}{ccc}
%   \centering
%   % Requires \usepackage{graphicx}
%  \includegraphics[width=7cm]{ratio}
%  \end{tabular}}
%   \caption{The $\Gamma_{B^\prime_1}$-dependence of $r(\Gamma_{B_1'})$ defined in Eq.~\ref{eq:retio} with $h$ unfixed (straight line) and $h$ fixed (dotted line).}\label{fig:ratio}
% \end{figure}

The coupling constants $g_2$ and $g_2'$ defined in Eq.~(\ref{Eq:LHS}) cannot be determined using the available data at present. Thus, one can not directly calculate the contributions from the $B_1^\prime \bar{B}^{(\ast)}$ and $B^{\ast}_0 \bar{B}^{\ast}$ meson loops. However, one can check the ratio of $\mathcal{B}(\Upsilon(5S) \to Z_b^+ \pi^-)$ and $\mathcal{B}(\Upsilon(5S)\to Z_b^{\prime +} \pi^-) $, which is independent of $g_2$. The estimated ratio is about 3. For the experimental data, we may take the values deduced from the row for the $\Upsilon(3S)$ in Table~\ref{Table:Exp}, for which the values of $\mathcal{B}_{Z_b}$ and $\mathcal{B}_{Z_b'}$ in the preliminary~\cite{Adachi:2012cx} and published~\cite{Garmash:2015rfd} Belle analyses are almost the same. This leads to a ratio about 1.8 up to a large uncertainty (it does not make much sense to give an uncertainty here from values in Table~\ref{Table:Exp} since we do not know the correlations). One may conclude that assuming that Fig.~\ref{fig:feyn-Zb-3} provides the dominant mechanism for the decays of the $\Upsilon(5S)\to Z_b^{(\prime)}\pi$, the ratio is roughly consistent with the data, and the value for $g_2$ is around 0.05~GeV$^{-1/2}$.

In Ref.~\cite{Abdesselam:2015zza}, the Belle Collaboration reported their energy scan measurements of  the $e^+e^- \to h_b(nP)\pi^+\pi^-$ $(n=1,2)$ cross sections, and the cross sections at around 10.999~GeV, the $\Upsilon(11020)$ mass, were fitted to be about $2.45$~pb and $4.05$~pb for $h_b(1P) \pi^+ \pi^-$ and $h_b(2P) \pi^+ \pi^-$, respectively. Assuming that the $h_b(nP) \pi^+ \pi^-$ are produced completely from the $\Upsilon(6S)$ at such an energy, and using the dilepton branching ratio of the $\Upsilon(6S)$, $\mathcal{B}(\Upsilon(6S) \to e^+ e^-)= (2.7^{+1.0}_{-0.8}) \times 10^{-6}$~\cite{Tanabashi:2018oca}, we can roughly estimate the branching ratios for $\Upsilon(6S) \to h_b(1P) \pi^+ \pi^-$ and $\Upsilon(6S) \to h_b(2P) \pi^+ \pi^-$ to be about $7.5\times 10^{-3}$ and $1.2 \times 10^{-2}$, respectively.
Assuming that Fig.~\ref{fig:feyn-Zb-3} provides the dominant mechanism for the decays of the $\Upsilon(6S)\to Z_b^{(\prime)}\pi$, our estimates of the branching ratios of the $\Upsilon(6S) \to Z_b^{(\prime) +} \pi^-$ are listed in Table~\ref{Table:BR}, depending on the unknown coupling constant $g_2^\prime$. The ratio of the so-obtained $\mathcal{B}(\Upsilon(6S) \to Z_b^{+} \pi^-)$ and $\mathcal{B}(\Upsilon(6S) \to Z_b^{\prime +} \pi^-)$ is about $2$.
Assuming that the $\Upsilon(6S) \to h_b(nP)\pi^+\pi^-$ proceed completely through the $Z_b$ and $Z_b'$ intermediate states, and using the measured branching ratios of $Z_b^{(\prime) } \to h_b(mP) \pi, \ (m=1,2)$ listed in Table~\ref{Table:Exp},  one can then roughly estimate the fractions of individual quasi-two-body contributions to $\Upsilon(6S)\to Z_b^{(\prime)\pm} \pi^\mp \to h_b(mP) \pi^+ \pi^- $, which are $f_{Z_b}= 46 \%, \ f_{Z_b^\prime}=54\%$ and $f_{Z_b}= 40 \%, \ f_{Z_b^\prime}=60\%$ for $h_b(1P) \pi^+ \pi^-$ and $h_b(2P) \pi^+ \pi^-$, respectively.
These so-predicted fractions are similar to those in the$\Upsilon(5S)$ case. With these predicted fractions and the branching ratios of $\Upsilon(6S)\to h_b(nP) \pi^+ \pi^-$ given above, one can estimate
\begin{eqnarray}
\mathcal{B}(\Upsilon(6S) \to Z_b^+  \pi^-) \sim 5\%\,, \nonumber\\
\mathcal{B}(\Upsilon(6S) \to Z_b^{\prime +}  \pi^-) \sim 3\%\,,
\end{eqnarray}
which could be tested in future measurements at Belle-II.
In addition, with these branching ratios and the results in Table~\ref{tab:state-2}, we get the coupling $g_2^\prime\sim0.05 \ \mathrm{GeV}^{-1/2}$, similar to the one for the $\Upsilon(5S)$.

Here, it should be noticed that the experimental data in Table.~\ref{Table:Exp} indicate strong $Z_b^{(\prime)} B^{(\ast)} B^{\ast}$ couplings, which is the basis of the meson-loop mechanism considered here. In the present estimation, all the involved coupling constants related to the $Z_b^{(\prime)}$ states are extracted from the corresponding experimental data, thus, one should get the same results regardless of the molecular or  tetraquark scenario for the $Z_b$ states.

\section{Summary}
\label{sec:summary}

Because the $Z_b$ states decay dominantly into the open-bottom final states, they must have strong couplings to the bottom-meson pairs. Thus the bottom-meson loops should be important for the production of the $Z_b$ states. Although the production rates from this kind of mechanism cannot be precisely predicted because of the lack of precise knowledge of the involved coupling constants, qualitative conclusions and rough estimates can be made.
In the present work, we investigate the contributions of the bottom-meson loops in the production of $Z_b^{(\prime)}$ from $\Upsilon(5S,6S)$ decays. Two kinds of bottom-meson loops connecting the initial bottomonia and the final $Z_b^{(\prime)} \pi$ are discussed, which are the $B^{(\ast)} \bar{B}^{(\ast)}+c.c.$ loops and the $B^\prime_1 \bar{B}^{(\ast)}/B^{\ast}_0 \bar{B}^{\ast} +c.c.$ loops. Using the NREFT power counting scheme, we argue that the latter one should dominate over the former. Such a conclusion is supported by numerical calculations assuming a natural value for the single unknown coupling constant in the latter case.

We then discuss the impact of the large widths of the $s_\ell^P=1/2^+$ bottom mesons, and point out that the large widths in fact help increase the importance of the $B^\prime_1 \bar{B}^{(\ast)}/B^{\ast}_0 \bar{B}^{\ast} +c.c.$ loops. The reason is that the widths are determined by the pionic coupling which also controls the magnitudes of the triangle diagrams explicitly.

Moreover, we present an estimate for the branching ratios of $\Upsilon(6S) \to Z_b \pi$ and $\Upsilon(6S) \to Z_b^{\prime} \pi$, which can be tested by future precise measurements at Belle-II.

\section*{ACKNOWLEDGMENTS}
This work is supported in part by the National Natural Science Foundation of China (NSFC) under Grants No.~11775050, No.~11375240, No.~11747601, and No.~11835015, by NSFC and Deutsche Forschungsgemeinschaft (DFG) through
funds provided to the Sino-German Collaborative Research Center ``Symmetries and the
Emergence of Structure in QCD'' (NSFC Grant No.~11621131001,
DFG Grant No.~TRR110), by the Thousand Talents Plan for Young
Professionals, by the CAS Key Research Program of Frontier Sciences
(Grant No.~QYZDB-SSW-SYS013), by the CAS Key Research Program (Grant No.~XDPB09), by the CAS Center for Excellence in Particle Physics (CCEPP) and by the Fundamental Research Funds for the Central Universities.

\begin{appendix}

\section{Decay amplitudes}\label{appendix-A}

Diagrams in Fig.~\ref{fig:feyn-Zb-1} indicate the $B^{(\ast)} \bar{B}^{(\ast)}+c.c.$ meson-loop contributions to $\Upsilon(5S)(p)\rightarrow Z_b^{(\prime)+}(p_z) \pi^-(q)$. The decay amplitude for the $\Upsilon(5S) \to Z_b^+ \pi^-$ reads
\begin{eqnarray}
\mathcal{M}_{a,b,c}&=&\frac{2\sqrt{2}z g_{1}g}{F_{\pi}}\{\vec{q}\cdot\vec{\epsilon}(p)\vec{q}\cdot\vec{\epsilon}(p_z)\times[-I^{(1)}(M_B, M_B, M_{B^\ast}, \vec{q})
\nonumber\\
&&+I^{(1)}(M_{B^\ast}, M_B, M_{B^\ast}, \vec{q})]+\vec{\epsilon}(P)\cdot\vec{\epsilon}(p_z)|\vec{q}|^2
\nonumber\\
&&\times[-I^{(1)}(M_{B^\ast}, M_B, M_{B^\ast}, \vec{q})+I^{(1)}(M_{B^\ast}, M_{B^\ast}, M_B, \vec{q})]\}.
\nonumber\\
\label{Eq:Amp1}
\end{eqnarray}

The decay amplitude for the $\Upsilon(5S) \to Z_b^{\prime +} \pi^-$ corresponds to Figs.~\ref{fig:feyn-Zb-1}(d)-\ref{fig:feyn-Zb-1}(e) reads
\begin{eqnarray}
\mathcal{M}_{d,e}&=&\frac{2\sqrt{2}z^\prime g_{1}g}{F_{\pi}}\{\vec{q}\cdot\vec{\epsilon}(p)\vec{q}\cdot\vec{\epsilon}(p_z)\times[-I^{(1)}(M_B,M_{B^\ast}, M_{B^\ast}, \vec{q})
\nonumber\\
&&+I^{(1)}(M_{B^\ast}, M_{B^\ast}, M_{B^\ast}, \vec{q})]+\vec{\epsilon}(P)\cdot\vec{\epsilon}(p_z)|\vec{q}|^2
\nonumber\\
&&\times[-I^{(1)}(M_B, M_{B^\ast}, M_{B^\ast}, \vec{q})+I^{(1)}(M_{B^\ast}, M_{B^\ast}, M_{B^\ast}, \vec{q})]\}.
\nonumber\\
\label{Eq:Amp2}
\end{eqnarray}

The $B^{(\ast)} B_1^{\prime}/B^\ast \bar{B}_0 +c.c$ loop diagrams are presented in Fig.~\ref{fig:feyn-Zb-3}. The decay amplitude
for the $\Upsilon(5S)(p)\to Z_b^+(p_z) \pi^-(q)$ corresponding to Figs.~\ref{fig:feyn-Zb-3}(a)-\ref{fig:feyn-Zb-3}(b) reads
\begin{eqnarray}
\bar{\mathcal{M}}_{a,b}&=&\frac{2\sqrt{2}zg_{2}h}{F_{\pi}}\epsilon^i(p)\epsilon^i(p_z)E_\pi
 I(M_{B^\prime_1}, M_B, M_{B^\ast}, \vec{q}) \nonumber\\
&&+\frac{2\sqrt{2}zg_{2}h}{F_{\pi}}\epsilon^i(p)\epsilon^i(p_z)E_\pi
 I(M_{B^\ast_0}, M_{B^\ast}, M_B , \vec{q}).~~~
 \label{Eq:Amp3}
\end{eqnarray}
The amplitude for the $\Upsilon(5S)(p)\to  Z^{\prime+}_b(p_z) \pi^-(q)$ corresponding to Fig.~\ref{fig:feyn-Zb-3}~(c) reads
\begin{eqnarray}
\bar{\mathcal{M}}_{c}&=&i\frac{4\sqrt{2}z^{\prime}g_2 h}{F_{\pi}}\epsilon^i(p)\epsilon^i(p_z)E_\pi I(M_{B^\prime_1},M_{B^\ast}, M_{B^\ast}, \vec{q}). \ \
\label{Eq:Amp4}
\end{eqnarray}

%\begin{widetext}
In above amplitudes, the basic three-point scalar loop function is defined as
{\small \begin{eqnarray}
&&I(m_1, m_2, m_3, \vec{q})\nonumber\\
 &=& i\int\frac{d^{4}l}{(2\pi)^4}\frac{1}{(l^2-m^2_1+i\epsilon)[(p-l)^2-m^2_2+i\epsilon][(l-q)^2-m^2_3+i\epsilon]}.\nonumber
 \end{eqnarray}}
One can work out an analytic expression for the above integral in the rest frame of the initial particle in the nonrelativistic approximation~\cite{Guo:2010ak}, which is,
\begin{eqnarray}
&&I(m_1, m_2, m_3, \vec{q})\nonumber\\
&\approx&\frac{\mu_{12}\mu_{23}}{16\pi m_1m_2m_3}\frac{1}{\sqrt{a}}\left[\tan^{-1}\frac{c_2-c_1}{2\sqrt{ac_1}}+\tan^{-1}\frac{2a+c_1-c_2}{2\sqrt{a(c_2-a)}}\right], \nonumber\\
\end{eqnarray}
where $\mu_{ij}=m_im_j/(m_i+m_j)$ are the reduced masses, $b_{12}=m_1+m_2-M$, $b_{23}=m_2+m_3+q^0-M$, and
\begin{equation}
 a=\left(\frac{\mu_{23}}{m_3}\right)^2\vec{q}^2, c_1=2\mu_{12}b_{12}, c_2=2\mu_{23}b_{23}+\frac{\mu_{23}}{m_3}\vec{q}^2.  \label{eq:c1c2}
\end{equation}
The involved vector loop integral in the rest frame of the initial particle is defined as
{\small
\begin{eqnarray}
&&q^iI^{(1)}(m_1, m_2, m_3, \vec{q})\nonumber\\&=&i\int\frac{d^{d}l}{(2\pi)^d}\frac{l^i}{(l^2-m^2_1+i\epsilon)[(p-l)^2-m^2_2+i\epsilon][(l-q)^2-m^2_3+i\epsilon]}.\nonumber\\
\end{eqnarray}
}
By using the technique of tensor reduction, we get the following nonrelativistic relation,
\begin{align}
&I^{(1)}(q)\approx\frac{\mu_{23}}{am_3}\left[B(c_2-a)-B(c)+\frac{1}{2}(c_2-c_1)I(q)\right],
\end{align}
where the function $B(c)$ is
\begin{align*}
&B(c)\equiv-\frac{\mu_{12}\mu_{23}}{4m_1m_2m_3}\frac{\sqrt{c-i\epsilon}}{4\pi}.
\end{align*}
%\end{widetext}

\end{appendix}

\end{document}